# Buffer Insertion for Bridges and Optimal Buffer Sizing for Communication Sub-System of Systems-on-Chip

Sankalp S. Kallakuri, Alex Doboli
ECE Department
Stony Brook University
Stony Brook , NY 11794
email: elsanky,adoboli@ece.sunysb.edu

Eugene A. Feinberg
AMS Department
Stony Brook University
Stony Brook , NY 11794
email: Eugene.Feinberg@stonybrook.edu

**Abstract**

*We have presented an optimal buffer sizing and buffer insertion methodology which uses stochastic models of the architecture and Continuous Time Markov Decision Processes CTMDPs. Such a methodology is useful in managing the scarce buffer resources available on chip as compared to network based data communication which can have large buffer space. The modeling of this problem in terms of a CTMDP framework lead to a nonlinear formulation due to usage of bridges in the bus architecture. We present a methodology to split the problem into several smaller though linear systems and we then solve these subsystems.*

## 1. Introduction

We have applied CTMDP (Continuous Time Markov Decision Processes) to optimise the buffer space used in SoC architectures. This involves using continuous time queueing models for the architectures. The use of such continuous time stochastic models is necessary due to the continuous time nature of tasks when they are executed on the IP cores and the shift from RTL level design to system level design. A finite amount of buffer space has to be distributed among a set of processors talking to a bus and the continuous time modeling allows incorporating how long certain amounts of buffer space have to be allotted as well as how much of the space should be allotted to processor. The division of the finite buffer space by certain stochastic policies generated through the CTMDP based solutions [1] could lead to an optimal division of the buffer resources. We found this optimal distribution of buffer space different from simple division of the space depending on traffic ratios.

While attempting to solve the buffer sizing problem we encountered a problem when there were bridges between buses. A typical example of such an architecture has been shown in Figure 1 where buses b,f and g talk to each other apart from processors. The architectures in which two buses are connected by a bridge which is a typical example in the AMBA and CoreConnect systems. For such a scenario with bridges the model developed for CTMDPs was nonlinear and the system of quadratic equations were not solvable for a test example shown in Figure 1. We solved this problem by splitting the system to smaller subsystems and solving linear equations obtained from CTMDP based methods for the subsystems.

CTMDPs and continuous time modeling have been used in work done by Pedram et al. and Marculescu et al. in the generation of power management policies. We have attempted to use similar stochastic modeling for optimal buffer sizing as well as distribution of finite buffer space.

## 2  Buffer Insertion with Split Subsystems

In Figure 1 the architecture has buses that are connected only to processors like bus a, as well as buses b,f and g which are connected to other buses too. Thus communication between processors 2,3 and 5 will involve insertion of buffers and will require the controller to take into account traffic from all three processors while making arbitration decisions for any of these three buses. One of the problems with designing such an arbiter is that it would require equations which would be quadratic in nature due to the interaction between two buses. Each bus by itself has been modeled by a linear set of equations. In case the buses talk to each other through bridges the equality constraints and the cost function have quadratic terms. The number of quadratic terms depend on how many points in the bus topology are there in which buses are connected to each other and an equation may have more than one quadratic term. An attempt was made to solve the nonlinear equations by using the nonlinear solver from Matlab ver. 6.1. but we were not able to get solutions for them.

The solution we propose for this problem is to split the bus architecture into a set of linear systems which are sep-arated from each other by buffers and solve a set of equations for each one of the independent linear modules. The system has been split into 4 subsystems as shown in Figures 2, each of which would have a set of linear equations associated with it. In order to find the optima for the entire system all the equations shall be solved in one go and not sequentially for each subsystem. In Figure 2 for subsystem 1 initially the buses b,f and g were communicating but after the split bus b becomes a shared resource between buffer b1, buffer b2, and processors 2 and 3 isolating it from buses f and g thus enabling us to write a set of linear equations for it. After solving the CTMDP for this system of equations and translating the state action pair probabilities into buffer space requirements by using the Kswitching policy [1] for a certain processor bus pair, the system is resimulated with the new buffer lengths and the losses are compared.

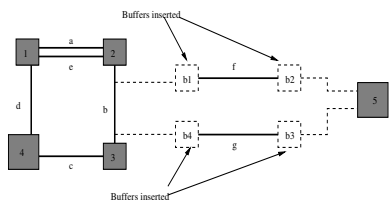

**Figure 1.** Sample Architecture

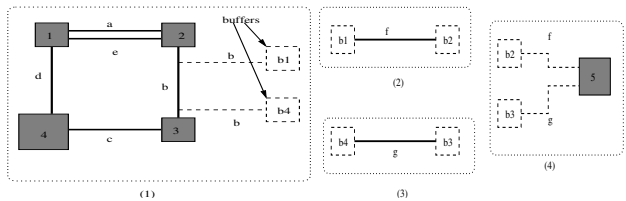

**Figure 2.** Split Subsystems

## 3 Experiments

The experiments used a network processor as a test architecture for the buffer insertion and buffer space distribution. In Figure 3 we have plotted the loss rates at the processors before and after the buffer sizing as the first and second bars of Figure 3. We found that though the loss rates decrease drastically for some processors for example processor 16 in Figure 3 they increase slightly for some processors for example processor 1. The third bar in Figure 3 are the loss rates for a timeout based policy, in which the processors request is not served if the data in the buffer times out i.e. reaches a threshold time. The threshold time chosen was the average time spent by a request in a buffer. We repeated these experiments for 10 iterations and found that though the loss may increase for some processors the overall loss of the system decreases by about 20% as compared to the constant buffer sizing policy and 50% for the timeout policy. We feel the difference before and after resizing could be improved with better profiling and weighing of the loss

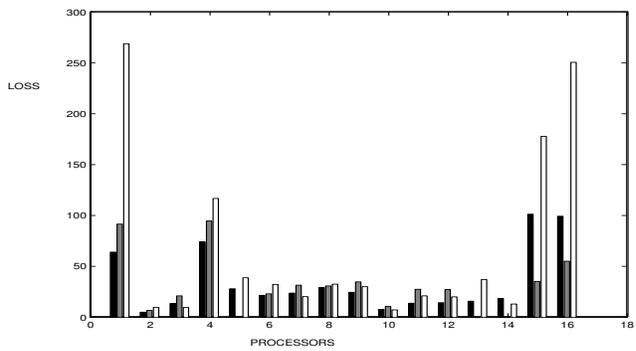

**Figure 3.** Loss rates before and after sizing

| PROCESSOR | Buf 160 | | Buf 320 | | Buf 640 | |
|---|---|---|---|---|---|---|
| | pre | post | pre | post | pre | post |
| 1 | 70 | 83 | 41 | 40 | 48 | 0 |
| 4 | 80 | 100 | 78 | 55 | 74 | 0 |
| 15 | 107 | 90 | 99 | 12 | 88 | 0 |
| 16 | 96 | 82 | 84 | 0 | 93 | 0 |

**Table 1.** Loss under varying total buffer size

at processors i.e. allowing some losses to be more important than the others.

In Table 1 we present the variation in the loss rates before and after sizing the buffers. We have presented the results only for a few processors which show significant variation but a similar trend was observed for the rest of the processors. We observed that some processors loss rates may increase when the buffer space is very limited as in the 160 units case and the redistribution doesn't provide much improvement as discussed in the previous paragraph. We increased the total buffer space from 160 to 320 and 640. The loss rates after resizing decreased with the increase in buffer space and fell to zero for the total buffer space of 640 units.

## 4 Conclusion

We have presented a methodology to judge the amount of buffer space required and in what points in the architecture it would be required by using stochastic models of the architecture. The use of CTMDP based methods gives us the optimal redistribution of the finite amount of buffer space so that loss is minimised, as seen in the experiments. The use of buffers for bridges can lead to efficient communication between two buses and buses can talk through them with reduced or no loss to other buses used by a different set of processors.